\documentclass[]{ecai} 

\usepackage{latexsym}
\usepackage{amssymb}
\usepackage{amsmath}
\usepackage{amsthm}
\usepackage{booktabs}
\usepackage{enumitem}
\usepackage{graphicx}
\usepackage{color}

\usepackage{hyperref}

\newtheorem{definition}{Definition}

\newcommand{\BibTeX}{B\kern-.05em{\sc i\kern-.025em b}\kern-.08em\TeX}

\begin{document}


\begin{frontmatter}


\paperid{99} 


\title{Fair Pairs: Fairness-Aware Ranking Recovery\\from Pairwise Comparisons}


\author[A]{\fnms{Georg}~\snm{Ahnert}\thanks{Corresponding Author. Email: ahnert [at] uni-mannheim [dot] de}}
\author[B,C,D]{\fnms{Antonio}~\snm{Ferrara}}
\author[C,E]{\fnms{Claudia}~\snm{Wagner}} 

\address[A]{University of Mannheim, Mannheim, Germany}
\address[B]{CENTAI, Turin, Italy}
\address[C]{GESIS -- Leibniz Institute for the Social Sciences, Cologne, Germany}
\address[D]{Graz University of Technology, Graz, Austria}
\address[E]{RWTH Aachen University, Aachen, Germany}


\begin{abstract}
Pairwise comparisons based on human judgements are an effective method for determining rankings of items or individuals. 
However, as human biases perpetuate from pairwise comparisons to recovered rankings, they affect algorithmic decision making.
In this paper, we introduce the problem of fairness-aware ranking recovery from pairwise comparisons.
We propose a group-conditioned accuracy measure which quantifies fairness of rankings recovered from pairwise comparisons.
We evaluate the impact of state-of-the-art ranking recovery algorithms and sampling approaches on accuracy and fairness of the recovered rankings, using synthetic and empirical data. 
Our results show that Fairness-Aware PageRank and GNNRank with FA*IR post-processing effectively mitigate existing biases in pairwise comparisons and improve the overall accuracy of recovered rankings. We highlight limitations and strengths of different approaches, and provide a Python package to facilitate replication and future work on fair ranking recovery from pairwise comparisons.
\end{abstract}

\end{frontmatter}


\section{Introduction} \label{sec:introduction}

Rankings of items or people are commonplace in, for instance, internet search~\cite{lerman2014leveraging}, academic admissions~\cite{waters2014grade}, hiring~\cite{geyik2019fairness}, or healthcare prioritization.
Recently, rankings recovered from pairwise comparisons have also been deployed for human alignment of Large Language Models~\cite{song2023preference}, to assess their outputs~\cite{li2023prd}, and to extract data from them~\cite{wu2023large}.
Pairwise comparisons, which are judgements between pairs of items or individuals, pose a viable type of data from which to generate rankings. They are less prone to inconsistencies~\cite{kiritchenko2017best} and judgement error~\cite{chen2013pairwise} than rating scale annotations.

\begin{figure}[t!]
    \vspace{-0.5cm}
    \centering
    \includegraphics[width=0.98\columnwidth]{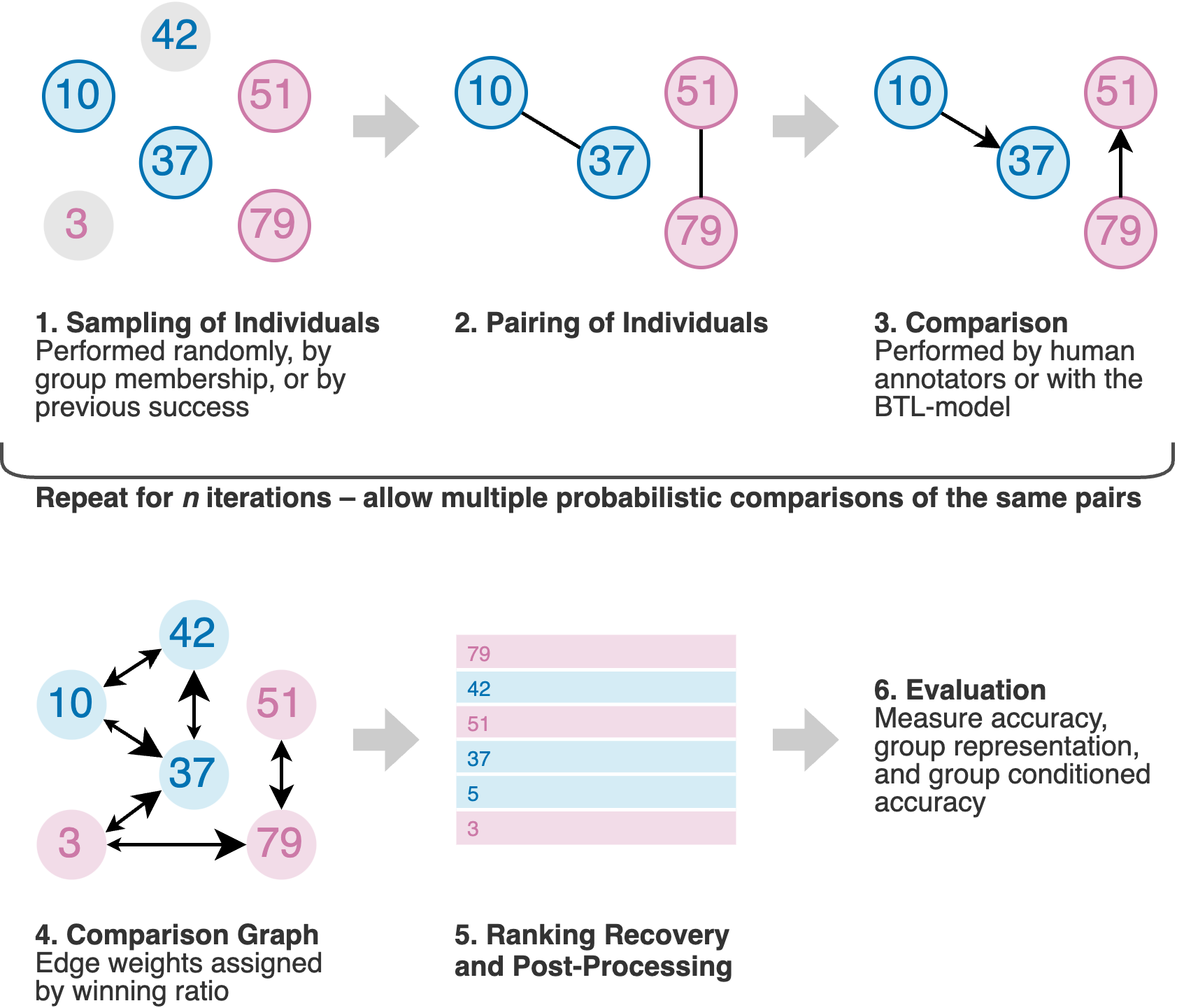}
    \vspace{0.1cm}
    \caption{\textbf{Research Setup} In this paper, we investigate the effect of sampling individuals for pairwise comparison and of ranking recovery methods, on fairness and accuracy of a recovered ranking.}
    \vspace{0.7cm}
    \label{fig:setup}
\end{figure}

However, pairwise comparisons can be subject to bias and systemic discrimination~\cite{mehrabi2021survey}. 
Research on algorithmic fairness has shown that with the increased relevance of machine learning, concerns regarding the introduction and perpetuation of discrimination must be taken seriously~\cite{chouldechova2020snapshot}.

While not without criticism~\cite{hagendorff2022blind}, research on the quantification of fairness lead to the development of fairness measures and the integration of fairness targets into machine learning algorithms~\cite{mehrabi2021survey}. In recent years, a sub-field on fairness in rankings has brought forth \mbox{pre-,} \mbox{in-,} and post-processing methods of ensuring fairness~\cite{zehlike2021fairness}. Still, to the best of our knowledge, existing ranking recovery methods have not been evaluated regarding their impact on fairness.

Our paper aims at bridging this gap by (i) applying a fairness-aware algorithm developed for a different task to the ranking recovery problem, (ii) introducing a group-conditioned error measure tailored towards ranking recovery, and (iii) evaluating the performance of state-of-the-art ranking recovery methods in terms of fairness and accuracy.

Our results on synthetic and empirical data show that under random sampling of individuals for comparison, resource-intense methods such as GNNRank~\cite{he2022gnnrank} have little benefit over heuristics-based approaches such as David's Score~\cite{david1987ranking}. We find that Fairness-Aware PageRank~\cite{tsioutsiouliklis2021fairness}, although not developed for ranking recovery from pairwise comparisons, is able to mitigate bias and to improve accuracy as measured against a latent ground-truth. Post-processing of recovered rankings using algorithms such as FA*IR~\cite{zehlike2017fa} can further improve overall accuracy at the expense of group-conditioned accuracy. We provide an open-source Python package alongside our paper to facilitate replication and future work on fair ranking recovery\footnote{\url{https://github.com/wanLo/fairpair}}.

\section{Related Work} \label{sec:related_work}
Measuring perceived attributes through pairwise comparisons dates back to \citet{thurstone1927law}, \citet{zermelo1929berechnung}, and \citet{kendall1940method}. Pairwise comparisons are generally incomplete and probabilistic, i.e., not all possible pairs are compared and a ``stronger'' individual might lose a comparison by chance. This probabilistic nature is also considered in models of rational choice theory~\cite{regenwetter2011transitivity}, and can lead to the formation of \textit{Condorcet cycles}, which are most prominently studied in the context of ballots~\cite{young1978consistent, gehrlein2010voting}. Only a limited number of comparisons between items or individuals can be drawn, pairing becomes crucial~\cite{ryvkin2005predictive}.

The context-agnostic problem of ranking recovery from pairwise comparisons has been addressed through approaches that can be roughly grouped into three categories: (i) heuristics-based methods such as David's Score~\cite{david1987ranking}, (ii) eigenvector- or random-walk-based such as RankCentrality~\cite{negahban2012iterative}, and (iii) similarity-based methods such as GNNRank~\cite{he2022gnnrank}. Considering all three categories, we include a diverse set of state-of-the-art algorithms into our evaluation and describe them in more detail in Section~\ref{subsec:ranking_recovery}.

Previous research evaluated the performance of ranking recovery methods regarding accuracy and efficiency. \citet{zhang2016crowdsourced} compare heuristics-based and learning-based ranking recovery methods. They conclude that ``local inference heuristics'' (e.g., David's Score) are outperformed by ``global inference heuristics'' (i.e., random walk based methods). Further, they show that ``global'' heuristics have comparable performance to learning-based methods, as is also shown by \citet{negahban2012iterative}. \citet{he2022gnnrank} include various baselines outperformed by GNNRank according to the authors' original error measure. The results indicate that similarity based methods might have an accuracy advantage over eigenvector-based methods. Yet, none of these evaluations takes bias or fairness into account.

Initial work on fairness in machine learning primarily focused on classification tasks~\cite{chouldechova2020snapshot}, with fairness in ranking recently attracting more attention~\cite{zehlike2021fairness}. Multiple methods have been proposed to achieve fair rankings in score-based ranking~\cite{yang2017measuring, celis2018ranking} and supervised learning-to-rank tasks~\cite{zehlike2017fa, singh2018fairness, geyik2019fairness, beutel2019fairness}. Both, however, substantially differ from ranking recovery from pairwise comparisons. In score-based ranking, a ranking is constructed given an utility score for each individual, and in supervised learning-to-rank, a ranking is predicted from the individuals' features. In contrast, pairwise comparisons between individuals is the only information available in ranking recovery from pairwise comparisons. Nevertheless, post-processing methods developed for score-based ranking and learning-to-rank tasks can also be applied to ranking recovery from pairwise comparisons. We include FA*IR~\cite{zehlike2017fa} and EPIRA~\cite{cachel2023fairer} into our evaluation.

Furthermore, it is worth to mention some works that deal with annotators' qualities and biases in ranking recovery from pairwise comparison methods \cite{ferrara2024bias, bugakova2019aggregation, chen2013pairwise}. However, in many real world scenarios annotators' information is not directly available. Differently, in our work, we focus on sampling strategies that affect the candidate selection in the pairwise comparisons and consequently their rankings.

\section{Approach} \label{sec:approach}
Our research setup is shown in Figure~\ref{fig:setup} and comprises six steps. We first establish pairwise comparisons through (step 1) sampling individuals from a population, (step 2) pairing them, and (step 3) comparing these pairs using human annotation or, when human annotation are not available, we simulate them with the Bradley-Terry-Luce (BTL) model~\cite{bradley1952rank}. Sampling, pairing and comparison is iteratively repeated to allow for multiple probabilistic comparisons of the same pair of individuals. We assume biases and discrimination to affect both the sampling and comparison steps, as we lay out in the following section, before describing derived sampling approaches.

From iterated pairwise comparisons, we obtain a comparison graph (step 4) in which edge weights indicate winning ratios. Ranking recovery methods take this comparison graph as input and output scores according to which the individuals in the graph are ranked (step 5). Post-processing might be applied to a recovered ranking to improve fairness. Finally, we evaluate the obtained ranking's error and fairness against the latent skill score (step 6). For group fairness, we consider both measures of group representation and group-conditioned accuracy.
We introduce the general task of ranking recovery from pairwise comparisons in Section~\ref{subsec:ranking_recovery}, as well as the algorithms we selected. 

\subsection{Biases and Discrimination} \label{subsec:biases}
We explore two stages in which bias and discrimination may enter and affect the final ranking: 
the selection of candidates for comparison and the comparison and ranking procedure.

\paragraph{Candidate selection:} Which individuals are included in pairwise comparisons can be subject to representation, popularity, or self-selection biases~\cite{mehrabi2021survey}. \textit{Representation bias} occurs when the individuals sampled are not representative of the population, e.g., women might be under-represented in pairwise comparisons of job performance in male-dominated fields. \textit{Popularity bias} describes popular items attracting more exposure, e.g., popular search results attract more pairwise comparisons as they appear at the top of the ranking. Finally, \textit{self-selection bias} can happen if individuals select themselves, e.g., students whose parents did not attend college might see themselves as ``not worthy'' of applying and thus will not be included in the comparison of scholarship candidates. To study the potential impact of these biases, we include three distinct sampling approaches into our research setup (Figure~\ref{fig:setup}, step 1): randomly, by group membership, or by previous success. These sampling approaches are laid out in more detail in the following section.

\begin{figure}[t!]
    \centering
    \includegraphics[width=\columnwidth]{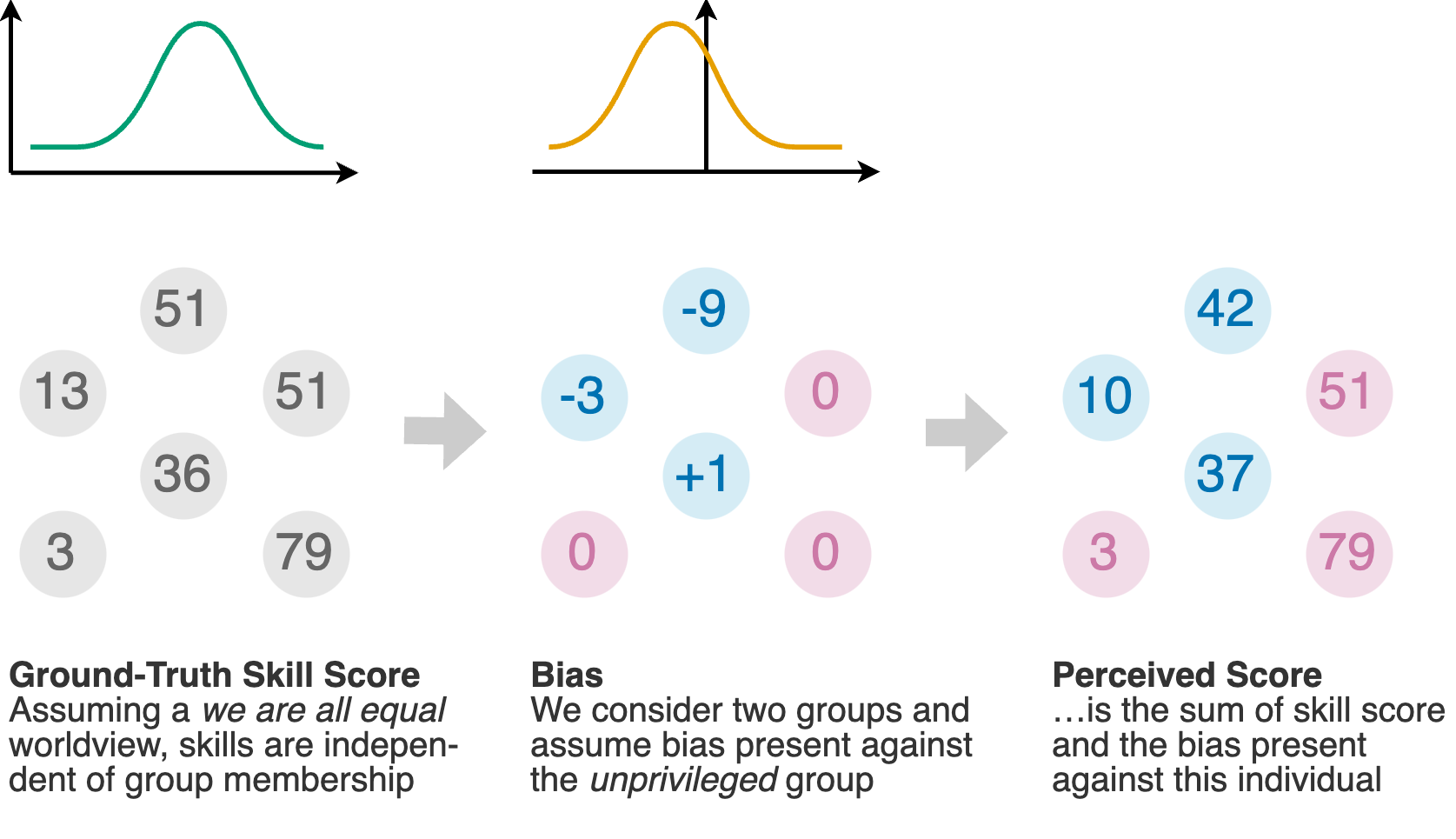}
    \vspace{0.1cm}
    \caption{\textbf{Normative Assumptions} Under the \textit{we are all equal} worldview~\cite{friedler2021possibility}, ground-truth \textit{skill score} and \textit{bias} are considered separately. If there are two groups, then the unprivileged group is subject to systemic discrimination, historical bias, or other types of biases. Ground-truth skill scores are independent of group membership, but perceived scores are impacted by group membership.}
    \label{fig:norm_assumptions}
    \vspace{0.7cm}
\end{figure}

\paragraph{Candidate comparison and ranking:} We consider systemic discrimination to be the origin of additional bias inherent to pairwise comparisons. \textit{Systemic discrimination} describes policies or customs that perpetuate discrimination, even if they are performed with no ill intent~\cite{mehrabi2021survey}. For instance, an individual from a poor family might lose pairwise comparisons of candidates for a managerial position by lacking the expected habitus. In our setup, we focus on a single, binary sensitive attribute that separates a population into a ``privileged'' group $G_\mathrm{priv}$ and an ``unprivileged'' group $G_\mathrm{unpriv}$ (Fig.~\ref{fig:norm_assumptions}). We assume a \textit{we are all equal} worldview~\cite{friedler2021possibility} in which each individual has a latent ground-truth \textit{skill score} on a given task, but pairwise comparisons are subject to \textit{bias}. This worldview is a necessary assumption for notions of group fairness~\cite{friedler2021possibility} which we apply in our research setup.

\subsection{Candidate Selection: Sampling Pairs} \label{subsec:sampling}

Resources for pairwise comparisons are generally limited, i.e., pairs must be selected for comparison. This can be seen as a two-step process, where individuals are (1) sampled from the population and then (2) paired, corresponding to steps 1 \& 2 in Figure~\ref{fig:setup}. We devise three distinct sampling approaches for step 1: Random Sampling, Oversampling, and Rank-Based Sampling. \textit{Random Sampling} serves as a baseline sampling approach where in each iteration, we select $20\%$ of individuals randomly from both groups. Since pairs are drawn without replacement, sampling already influences the possible pairings. For instance, if oversampling is applied, then there will always be different numbers of homogeneous pairs (both individuals from the same group) among the groups. Without oversampling, the number of same and different groups pairs is (almost) the same. We fix pairing to be random since biases from the literature~\cite{mehrabi2021survey} are more closely linked to sampling approaches rather than pairing.

\paragraph{Oversampling.} One aspect according to which individuals could be sampled is group-membership. If designers of a pairwise competition are aware of \textit{representation bias} (i.e., a group being under-represented in a sample), they might stratify their sample by over-sampling this group. Further, designers of a pairwise competition might try to mitigate systemic discrimination or historical bias by allotting an unprivileged group disproportionate chances to compete (i.e., applying a form of affirmative action). For instance, this could be desirable for female job applicants in a male-dominated field. We therefore implement \textit{Oversampling} as a group-dependent sampling approach, sampling more individuals from the unprivileged group. In this paper, we fix the oversampling rate at $75\%$, i.e. we sample three times more individuals from the unprivileged group. We investigated different oversampling rates in a preliminary study and include the results in Appendix C.5 in the Supplementary Material. The effects of oversampling (especially for David's Score and RankCentrality) increase with an increasing oversampling rate. 

\paragraph{Rank-Based Sampling.} Another possible aspect for sampling individuals is previous success. Our iterative setup of sampling, pairing, and comparing allows for the consideration of feedback loops between these components. Both \textit{popularity} and \textit{self-selection bias} can be the result of such a feedback loop, as they give more successful individuals a higher chance of being selected again for comparison. On one hand, popularity bias might be desirable if the accuracy of the top of the ranking matters most. On the other hand, the Matthew-Effect~\cite{merton1988matthew} can be a type of self-selection bias if more successful individuals are more eager to compete again. We thus implement \textit{Rank-Based Sampling} using intermittently recovered ranks to condition the probabilities for sampling nodes in the next iteration. We normalize these probabilities to be independent of ranking length, to give exponentially higher chances to top-ranking individuals, and to keep a minimal probability for selection even at the bottom of the ranking.

\subsection{Candidate Comparison and Ranking Recovery} \label{subsec:ranking_recovery}

We are given a population $N = \{1,\dots,n\}$ and two subgroups $G_\mathrm{priv}, G_\mathrm{unpriv} \subseteq N$. We assume that each individual $i \in N$ has a ground truth unobservable latent skill score $t_i$. The latent skill scores also induce a ground truth unobservable ranking $r^*$. Ranking recovery methods aim to reconstruct the latent skill scores $t$ and hence recover the ranking $r^*$. Furthermore, we assume that each item has a perceived score $s_i$, with the difference between latent skill score and perceived score called bias. While we cannot observe the perceived scores $s$ either, they inform the human judgments and probabilistically determine the pairwise comparisons. 

From the human evaluated pairwise comparisons, it is possible to construct a pairwise comparisons graph, that is exploited by various ranking recovery methods, as follows. Let the individuals be the nodes of a directed, weighted comparison graph. For each pairwise comparison we observe between distinct individuals $i$ and $j$, we then update the weight of the edge $(i,j)$ to reflect the proportion of $j$'s wins against $i$ over the total number of comparisons between the pair. This results in an adjacency matrix $A$, with zeros on the diagonal (i.e., no self-loops), and in which $A_{ij} = 1 - A_{ji}$ for each pair $i,j$ that has been compared at least once. We assume the pairwise comparisons to be incomplete and potentially inconsistent. In other words, there are individuals $i,j$ for which $A_{ij} = A_{ji} = 0$, and the comparison graph contains Condorcet cycles such as $A_{ij} = A_{jk} = A_{ki} = 1$.

Existing ranking recovery algorithms are based on heuristics, random walks, or similarity, and we include one state-of-the-art algorithm from each class into our evaluation. 
As a baseline, we also include a \textit{Random Rank Recovery} method that randomly assigns ranks to candidates.

\paragraph{David's Score.}
David's Score~\cite{david1987ranking} is a heuristics-based approach to ranking recovery that estimates an individual $i$'s rank from its direct wins and losses as follows: $\mathrm{DS}_i = w_i + \bar{w_i} - l_i - \bar{l_i}$, where $w_i = \sum_j A_{ji}$ is the sum of its winning ratios and $l_i = \sum_j A_{ij}$ the sum of its losing ratios. $\bar{w_i} = \sum_j A_{ji} w_j$ and $\bar{l_i} = \sum_j A_{ij} l_j$ aggregate these sums from individuals $j$ that $i$ is compared to. While David's Score does not require a connected comparison graph and can be computed very efficiently, it only considers an individual's success (i.e., proportion of wins) over its neighbors and their successes. However, there might be a community of individuals that are weaker than everyone else globally, in which one individual would still have a high proportion of wins over the other individuals in this community locally.

\paragraph{RankCentrality.}
A possible ``global heuristic'' for ranking recovery is the stationary distribution of a random walk over the comparison graph. Using this idea, RankCentrality~\cite{negahban2012iterative} obtains a ranking as follows: First, the adjacency matrix is normalized through dividing by the individuals' out-degrees and adding self-loops. Then, the stationary distribution of a random walk is calculated as the top left eigenvector of this normalized adjacency matrix.

\paragraph{GNNRank.}
Another approach to ranking recovery assumes that individuals who win similarly often against the same other individuals should be ranked similarly. SerialRank~\cite{fogel2014serialrank} uses the Fiedler vector, i.e., the second smallest eigenvector of the Laplacian of the similarity matrix, to calculate this similarity. SerialRank does, however, require a considerable amount of consistent comparisons to recover an accurate ranking. GNNRank~\cite{he2022gnnrank} tries to overcome this limitation by learning better suited similarity vectors using node embeddings obtained from the comparison graph through the use of graph neural networks.

\paragraph{Fairness-Aware PageRank.}
The PageRank algorithm~\cite{page1999pagerank} is designed to estimate the relevance of web pages based on the hyperlink network in which they are embedded. Using a random walk with restarts, it solves a problem related to, yet distinct from ranking recovery from pairwise comparisons. Fairness-Aware PageRank~\cite{tsioutsiouliklis2021fairness} introduces group-awareness into PageRank by equally distributing PageRank mass (i.e., relevance) to all subgroups.
Fairness-Aware PageRank is similar to RankCentrality in that it belongs to the group of random walk based approaches to network centrality.
While Fairness-Aware PageRank neglects edge weights, it allows for multi-edges in the comparison graph, i.e., individual comparisons can be expressed separately. In addition to a winning ratio, multi-edges express the total number of comparisons between a pair. This is a potential benefit of Fairness-Aware PageRank over methods that allow for edge weights but not for multi-edges.

\paragraph{Post-Processing.}
In addition to the discussed ranking recovery methods, we also included two post-processing methods into our evaluation. FA*IR~\cite{zehlike2017fa} is a commonly used post-processing method that re-ranks individuals to ensure a specified representation in the top-k of a ranking, while maximizing utility. EPIRA~\cite{cachel2023fairer} is a post-processing method tailored towards the exposure measure of group representation, which we also employ in our setup.

\subsection{Measures} \label{subsec:measures}

Given our research setup, a suitable error measure should (i) compare the recovered ranking against the ground truth skill scores, (ii) assign higher penalties to gross differences between order of skill scores and ranks, and (iii) consider subgroups in the ranking. We modify the normalization term of \citet{negahban2012iterative}'s Weighted Kemeny Distance to also satisfy the third desired property.
Kemeny Distance~\cite{kemeny1959mathematics} counts discordant pairs between the recovered ranking and the ground-truth skill scores, but requires normalization to make it comparable between rankings of different lengths. Weighted Kemeny Distance~\cite{negahban2012iterative} introduces more sensitivity to discordant pairs that have a large weight difference and is normalized with regard to the number of nodes and the norm of the weight vector. This normalization cannot be applied to find a subgroup's error, which we would expect to comprise both within-group and between-groups comparisons. We introduce Group-Conditioned Weighted Kemeny Distance by instead normalizing on all pairs involving a group, i.e., within- \& between-group pairs. This normalization guarantees the error to be in the interval $ [0,1] \subset \mathbb{R}$ and to be comparable between subgroups of different size.

\begin{definition}[Group-Conditioned Weighted Kemeny Distance]
    \label{def:groupWeightedKemeny}
    For a subgroup $G$ of a population $N$ ($G \subseteq N$) with ground-truth \textit{skill scores} $t$ and recovered ranks $r$, we define the Group-Conditioned Weighted Kemeny Distance as:
    \begin{align}
        D^G_t(r) := \sqrt{\frac{\sum_{i<j} (t_i-t_j)^2 I_{(t_i-t_j)(r_i-r_j)>0}}{\sum_{i<j} (t_i-t_j)^2}}
    \end{align}
    for all pairs of individuals $(i,j)$ such that $i\in N$ and $j\in G$, comparing the individuals of a group both in-group and out-of-group for evaluation. Further, $I_{[\ ]}$ is the indicator function that picks discordant pairs. Given a privileged group $G_\mathrm{priv}$ and an unprivileged group $G_\mathrm{unpriv}$, we define the \textbf{error difference} as:
    \begin{align}
        D^\mathrm{diff}_t(r) = D^{G_\mathrm{unpriv}}_t(r) - D^{G_\mathrm{priv}}_t(r)
    \end{align}
\end{definition}

In this paper, we consider both the overall error of a recovered ranking and the error difference as a measure of fairness. Furthermore, we include \textit{exposure} as a measure of group representation as introduced by \citet{singh2018fairness}. In contrast to the (group-conditioned) error measure, exposure only takes to group membership of individuals into account, not their ground-truth skill scores. A logarithmic discount then ensures a focus on the top-k individual's groups.

\begin{definition}[Exposure]
    \label{def:exposure}
    Group representation as exposure is measured for a group $G$ in a ranking $r$ as:
    \begin{align}
        \mathrm{Exp}^G(r):=\frac{1}{|G|}\sum_{d\in G}\frac{1}{\log_2(r_d + 2)}
    \end{align}

    Given a privileged group $G_\mathrm{priv}$ and an unprivileged group $G_\mathrm{unpriv}$, we define the \textbf{exposure difference} as:
    \begin{align}
        \mathrm{Exp}^\mathrm{diff}(r) = \mathrm{Exp}^{G_\mathrm{unpriv}}(r) - \mathrm{Exp}^{G_\mathrm{priv}}(r)
    \end{align}
\end{definition}

\begin{figure}[t!]
    \centering
    \includegraphics[width=\columnwidth]{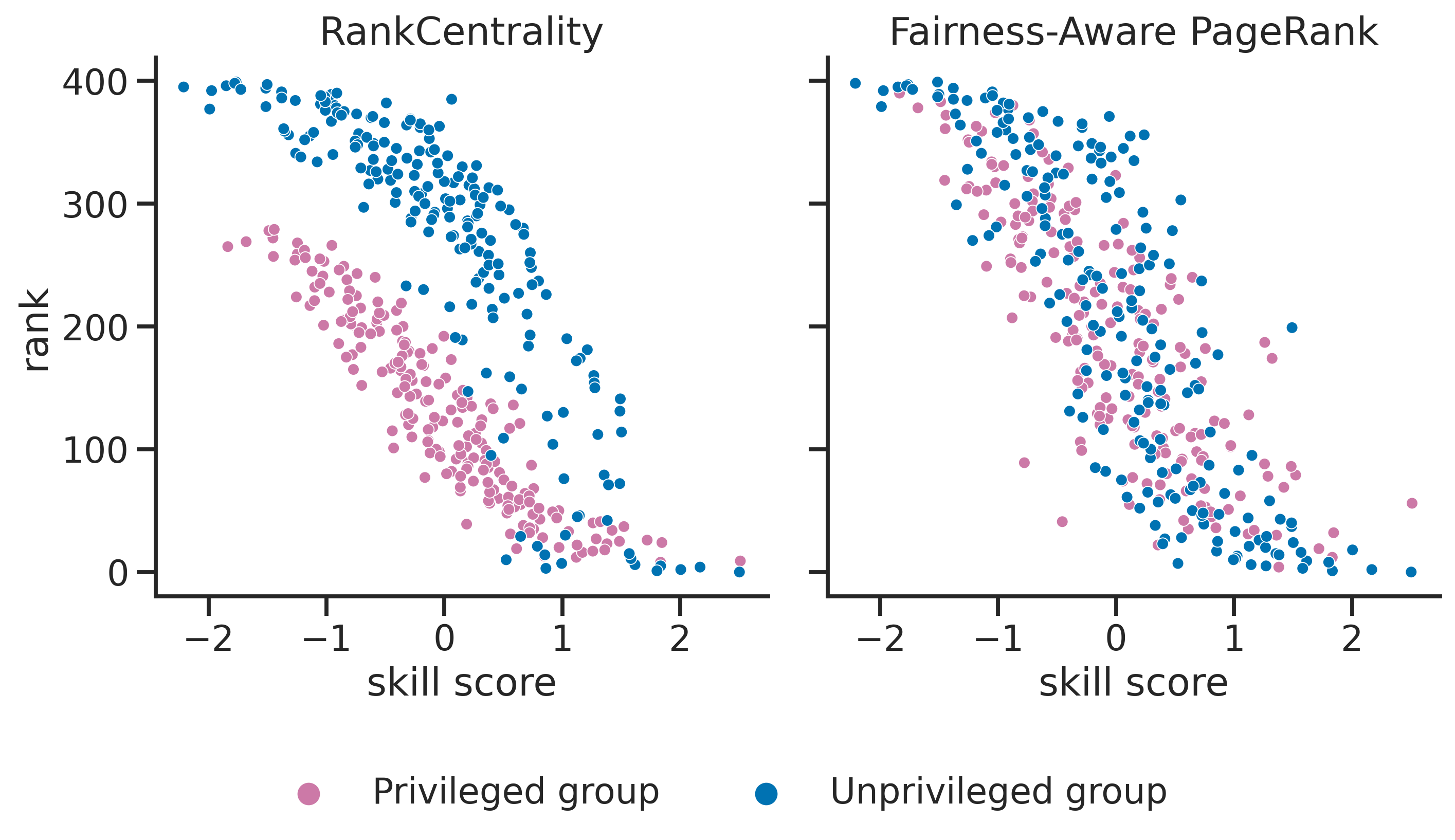}
    \vspace{0.1cm}
    \caption{\textbf{Correlations of Skill Score (higher is better) and Rank (lower is better) by Recovery Method.} 400 individuals in 2 equal size groups, compared in 1000 iterations using Oversampling and the BTL model. \textbf{Left:} Ranks recovered with RankCentrality~\cite{negahban2012iterative}. Unprivileged individuals have a higher mean rank because of bias, but are sorted to the extremes of the ranking when Oversampling is applied with RankCentrality. \textbf{Right:} Ranks recovered with Fairness-Aware PageRank~\cite{tsioutsiouliklis2021fairness}. The correlations of both groups overlap, but within-group error is higher.}
    \label{fig:correlations}
    \vspace{0.7cm}
\end{figure}

\section{Experiments} \label{sec:experiments}

Given the introduced sampling strategies and ranking recovery methods, as well as the setup illustrated in Figure~\ref{fig:setup}, we conduct experiments on both synthetic and empirical data.

\subsection{Synthetic Data} \label{subsec:synthetic_data}

\begin{figure*}[ht!]
    \centering
    \includegraphics[width=\textwidth]{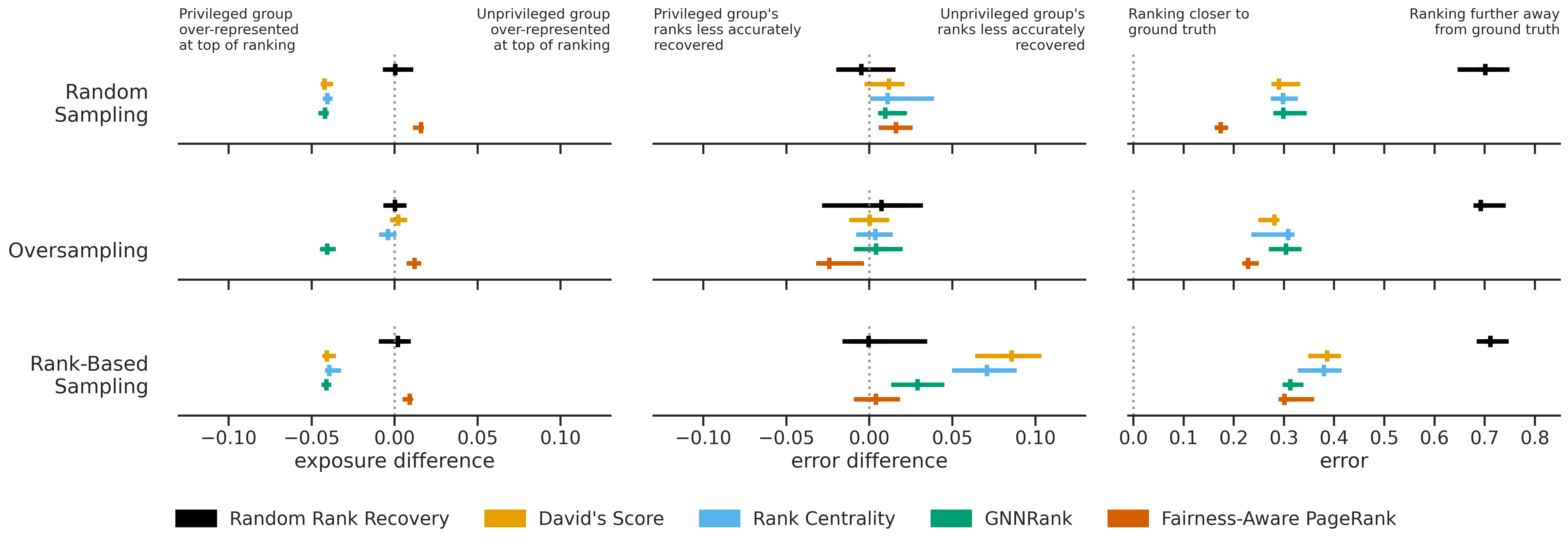}
    \vspace{0.1cm}
    \caption{\textbf{Results after 500 Iterations of Simulated Pairwise Comparisons, by Sampling Approach and Ranking Recovery Method.} Medians and ranges of 10 trials. Exposure difference (left) and error difference (center) are group-conditioned measures of fairness, error (right) reflects the whole ranking -- dashed lines indicate optimal values. David's Score and RankCentrality intermittently remove exposure difference when Oversampling is applied. Error difference is minimized by GNNRank and Fairness-Aware PageRank, even under Rank-Based Sampling. Fairness-Aware PageRank over-shoots on the unprivileged group's exposure but achieves best overall accuracy by partially mitigating bias. }
    \label{fig:sim_results_no_postprocessing}
    \vspace{0.7cm}
\end{figure*}

We consider a synthetic dataset of 400 individuals divided in two equal size groups. First, we draw \textit{skill scores} from the same normal distribution $\mathcal{N_\mathrm{skill}}(\mu_\mathrm{skill}, \sigma_\mathrm{skill})$ for all individuals, then we add \textit{bias} drawn from a different normal distribution $\mathcal{N_\mathrm{bias}}(\mu_\mathrm{bias}, \sigma_\mathrm{bias})$ to the scores of the unprivileged group. For results on a dataset in which \textit{skill scores} and \textit{bias} are not normally distributed, see Section~\ref{subsec:empirical_data}. For the privileged group, \textit{skill scores} and \textit{perceived scores} are identical, while for the unprivileged group, \textit{perceived scores} comprise \textit{bias}. This is without loss of generality since what is relevant in the pairwise comparisons is the relative difference in the (perceived) scores.
We use the Bradley-Terry-Luce (BTL) model~\cite{bradley1952rank} to simulate human comparisons.
The magnitude of scores matters for the BTL model, as it determines whether an individual is almost certain to win a comparison or only slightly more likely to win than to loose a comparison.
To inform our choice of parameters for the \textit{skill score} and \textit{bias} distributions, we propose fixing the expected probabilities of (i) stronger individuals winning ($p_\mathrm{stronger}$), and (ii) privileged individuals winning ($p_\mathrm{discr}$) a pairwise comparison. We explain how to obtain distribution parameters from these probabilities in detail in Appendix A in the Supplementary Material.

To simulate average amounts of comparison uncertainty and of bias against the unprivileged group, we set $p_\mathrm{stronger} = p_\mathrm{discr} = 75\%$, as well as $\mu_\mathrm{skill} = 0$ and $\sigma_\mathrm{bias} = \sigma_\mathrm{skill}/2$. In general, both the type of distribution and the parameters should be calibrated to a specific context. The IMDB-WIKI-SbS dataset discussed in Section~\ref{subsec:empirical_data}, for instance, shows more conservative probabilities of $p_\mathrm{stronger} = 81.6\%$ and $p_\mathrm{discr} = 62.7\%$.

We iteratively select $20\%$ of all individuals according to the sampling approaches introduced in Section~\ref{subsec:sampling}, pair them randomly, and probabilistically compare the pairs using the BTL model~\cite{bradley1952rank}, also known as the \textit{softmax} function.

\begin{definition}[The Bradley-Terry-Luce (BTL) Model]
    \label{def:expBTLmodel}
    For two individuals $i$ and $j$ with \textit{perceived scores} $s_i$ and $s_j$, the probability of $i$ winning a pairwise comparison against $j$ is given by:
    \begin{align}
        \mathrm{P}(i\text{ wins over } j) = \frac{e^{s_i}}{e^{s_i} + e^{s_j}} = \frac{1}{1 + e^{s_j - s_i}}
    \end{align}
\end{definition}

Given the probability, we randomly identify the winner of a comparison. If we re-compare the same individuals $i$ and $j$ multiple times, we update the edge weight from $j$ to $i$ with the winning ratio of $j$ over $i$ and vice versa.

\subsection{Results from Simulated Comparisons} \label{subsec:simulated_results}

We simulate up to 1000 iterations for each combination of sampling approach and ranking recovery method. Two different outcomes of such a ranking recovery are illustrated in Figure~\ref{fig:correlations}. \textit{Exposure difference}, \textit{error difference}, and \textit{error} (medians and ranges of 10 trials) are shown after 500 iterations in Figure~\ref{fig:sim_results_no_postprocessing}. For detailed results per iteration see Appendix C.1 in the Supplementary Material.

Regarding \textit{exposure difference}, we find that David's Score and RankCentrality in combination with Oversampling indeed lead to improved exposure of the unprivileged group. Upon closer inspection (Fig.~\ref{fig:correlations}), this is because RankCentrality sorts the oversampled group to both extremes of the ranking and can therefore not be relied upon to remedy existing bias. Whether this effect temporarily mitigates exposure differences as shown in Figure~\ref{fig:sim_results_no_postprocessing} depends on the amount of bias present in the data, so the oversampling rate might need to be adjusted accordingly. Still, we recommend against relying on this effect to mitigate bias, as it diminishes with an increasing number of comparisons being drawn. Fairness-Aware PageRank, by contrast, consistently improves the exposure of the unprivileged group across all sampling strategies and iterations. This ranking recovery method does, however, slightly overshoot and over-represents the unprivileged group at the top of the ranking.

Random Sampling and Oversampling have very little effect on \textit{error difference}. Only Rank-Based Sampling creates higher error in the ranks recovered for the unprivileged group. This comes as no surprise, given that the privileged individuals are more likely to be found at the top of the ranking and will thus be compared more often. Among the ranking recovery methods that are not group-aware, GNNRank deals best with the Rank-Based Sampling scenario in terms of \textit{error difference} and overall \textit{error}.

\begin{figure*}[ht!]
    \centering
    \includegraphics[width=\textwidth]{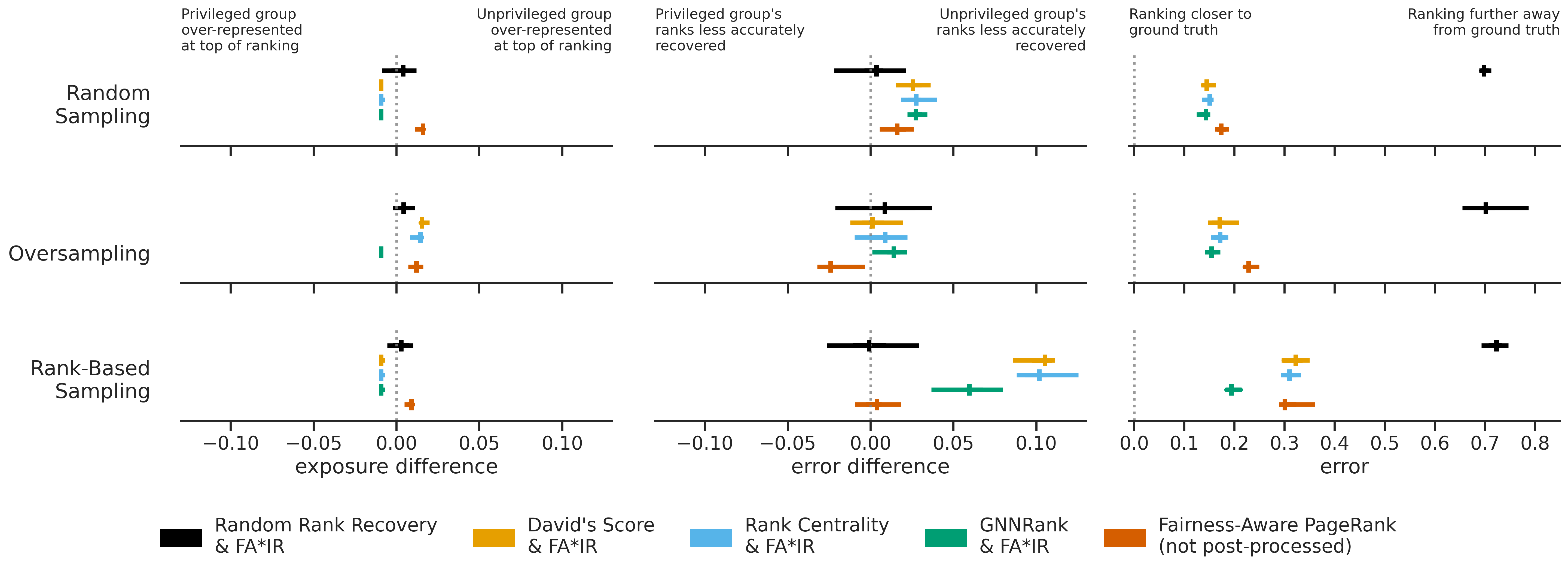}
    \vspace{0.1cm}
    \caption{\textbf{Post-Processed Results after 500 Iterations of Simulated Pairwise Comparisons, by Sampling Approach and Ranking Recovery Method.} Post-processing was performed using the FA*IR algorithm~\cite{zehlike2017fa} with $p=0.6$ and $\alpha = 0.1$. FA*IR is able to effectively limit exposure difference while not over-shooting the way Fairness-Aware PageRank does. The post-processing technique also improves overall accuracy and is able to outperform Fairness-Aware PageRank, in particular if paired with GNNRank for ranking recovery. FA*IR does, in contrast to Fairness-Aware PageRank, negatively impact error difference.}
    \label{fig:sim_results_FAstarIRp60}
    \vspace{0.7cm}
\end{figure*}

In fact, if we do not simulate bias, GNNRank consistently performs similar to or outperforms David's Score and RankCentrality regarding \textit{error}. This replicates the findings by~\citet{he2022gnnrank} but using a different error measure.
Notwithstanding GNNRank's performance in scenarios without bias, Fairness-Aware PageRank consistently outperforms the other ranking recovery methods if bias is assumed to be present. This is because it is the only group-aware recovery method investigated and it shares a \textit{we are all equal} worldview with our research setup. Further, Fairness-Aware PageRank is capable of minimizing error difference in Rank-Based Sampling scenarios as well.

\paragraph{Post-Processing Results.}
We apply the EPIRA~\cite{cachel2023fairer} post-processing method with the $\mathrm{bnd}$ parameter set to both $0.9$ and $0.99$. The EPIRA method optimizes a ranking until $\mathrm{Exp}^{G_\mathrm{unpriv}}(r) / \mathrm{Exp}^{G_\mathrm{priv}}(r) \geq \mathrm{bnd}$. We find this method to effectively mitigate exposure difference. However, post-processing with EPIRA does not improve error or error difference, as shown in Appendix C.2. This is what we would expect from a post-processing method that optimizes this very measure (i.e., exposure) with minimal swaps in the ranking.

The FA*IR post-processing method~\cite{zehlike2017fa} has a larger impact on all individuals in the ranking. We find that it effectively reduces exposure difference while not overshooting and also greatly reduces error. Setting the proportion of protected individuals $p$ to a value even larger than $0.5$ yields optimal results in our setup (see Fig.~\ref{fig:sim_results_FAstarIRp60}). In combination with GNNRank, FA*IR is able to outperform Fairness-Aware PageRank regarding overall error, in particular under Rank-Based Sampling. This combination of ranking recovery and post-processing methods does, however, not mitigate the error difference observable under this sampling approach. Thus, Fairness-Aware PageRank might still be preferable.\footnote{See also Appendix C.3 in the Supplementary Material for more details.}

\subsection{Empirical Data} \label{subsec:empirical_data}

Validating our findings on empirical data necessitates a dataset where both ground-truth values (i.e., \textit{skill scores}) and pairwise comparisons are available and \textit{bias} is present in the data. The IMDB-WIKI-SbS dataset~\cite{pavlichenko2021imdb} is a recently released large-scale dataset consisting of human-annotated pairwise comparisons of age between faces from IMDB.com and Wikipedia. The dataset consists in
9,150 images sampled from the IMDB-WIKI dataset~\cite{rothe2018imdb} that are compared in 250,249 pairs by crowdworkers, tasked with selecting the older individual of each pair. 

\citet{pavlichenko2021imdb}, however, already acknowledge quality issues in the dataset concerning the ground-truth labels for age and gender.
To overcome these issues, we applied FairFace~\cite{karkkainen2021fairface} to all pictures and classified age and gender. We then retrieved the image captions for all original images from IMDB.com and extracted a list of linked actors for each image. Using the actor's IMDB IDs, we extracted their date of birth and gender from Wikidata and where able to reconstruct their approximate age given the year that the picture was taken. We then performed an exact match on gender and a closest match on age between each actor and the detected gender and age for a given face. We further removed all pictures in which no age match could be established within the age range identified by FairFace $\pm 5$ years.

\begin{figure*}[t!]
    \centering
    \includegraphics[width=\textwidth]{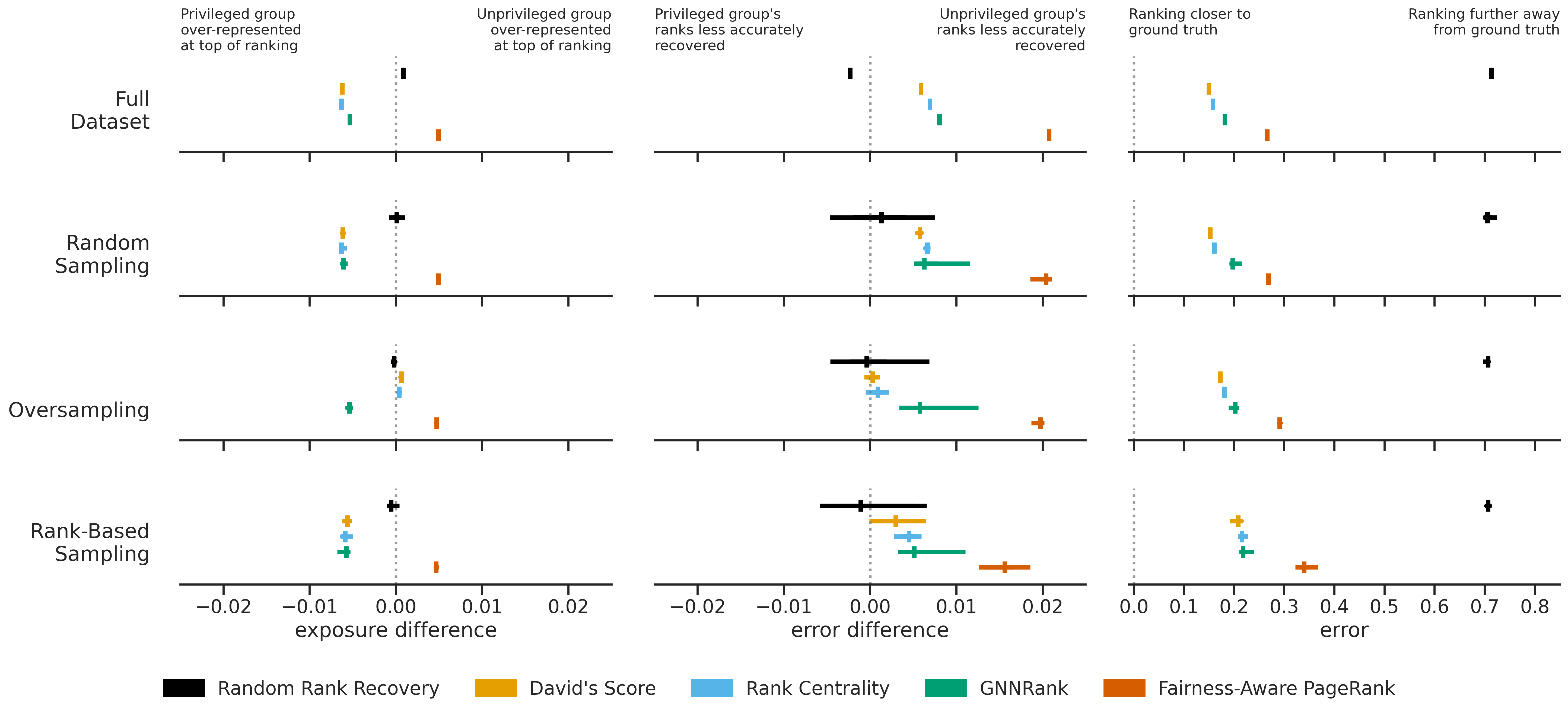}
    \vspace{0.1cm}
    \caption{\textbf{Results from the IMDB-WIKI-SbS dataset~\cite{pavlichenko2021imdb} with improved labels, by Sub-Sampling Approach and Ranking Recovery Method.} Ranks were either recovered from the full dataset, or after 500 iterations of sub-sampling with a specified sampling approach. Similar to the simulated comparisons data, Fairness-Aware PageRank overshoots on improving the unprivileged group's exposure. Further, Oversampling in combination with David's Score or RankCentrality leads to improved exposure for the unprivileged group as well. In contrast to our simulations, error is higher overall and for the unprivileged group when Fairness-Aware PageRank is applied in contrast to other ranking recovery methods. This can be explained by less bias and a less dense comparison graph observed for the IMDB-WIKI-SbS dataset.}
    \label{fig:imdb_results_no_postprocessing}
    \vspace{0.7cm}
\end{figure*}

Since pairwise comparisons are already given, we modify our sampling strategies to sub-sample from the existing comparison graph. We follow the same objectives when sampling (i.e., random/group-based/success-based) and then choose among existing edges instead of randomly pairing the sampled individuals.
After re-labeling, we have 6,123 images of 3,430 men and 2,693 women (i.e., unequal group size). Women are on average younger than man (i.e., unequal \textit{skill score} distributions). Because of the domains from which the pictures where gathered (movies and Wikipedia) our data is skewed towards young people, i.e., age does not follow a normal distribution. Our results, as shown in Figure~\ref{fig:imdb_results_no_postprocessing}, finally also indicate the existence of \textit{bias} in the form of gender ageism, i.e., women being systematically perceived younger than men of equal age.

\subsection{Empirical Results} \label{subsec:empirical_results}

We observe the following similarities to results from synthetic data. Firstly, David's Score and RankCentrality partially mitigate the under-representation of the unprivileged group under Oversampling. As observed with synthetic data, however, this effect cannot be relied upon. Next, Fairness-Aware PageRank overshoots in correcting for the under-representation of the unprivileged group.

We found that GNNRank generalized less well on the IMDB-WIKI-SbS comparison graph and thus re-trained the model after each iteration of sampling and comparison. In contrast to our experiments on synthetic data, GNNRank shows no advantage in error over the other ranking recovery methods, even under Rank-Based Sampling. We suspect this to be the case because of the absence of edge weights for the comparisons, i.e., each pair was only compared once. With less graph structure available, GNNRank cannot find better node embeddings, even if it is re-trained separately after each iteration.

Finally, Fairness-Aware PageRank performs worse than the other ranking recovery methods on the empirical dataset regarding both overall error and error difference. Our explanation for this result is twofold. Firstly, the observed bias against women is much smaller than in our simulations and it thus becomes more apparent that Fairness-Aware PageRank is designed to solve a different task. Secondly, Fairness-Aware PageRank assumes the same ground-truth age distribution for both genders and therefore recovers a non-optimal ranking from the IMDB-WIKI-SbS comparison graph in which this assumption is not met.

The FA*IR post-processing algorithm cannot be applied to a ranking recovered from the IMDB-WIKI-SbS dataset, since only rankings with at most 400 individuals are supported and running the Python implementation\footnote{\url{https://github.com/fair-search/fairsearch-fair-python}} nonetheless results in a \texttt{RecursionError}. Applying EPIRA to the rankings recovered from IMDB-WIKI-SbS has little impact, as shown in Appendix C.4 in the Supplementary Material.

\section{Conclusion} \label{sec:conclusion}

In this paper, we extend previous research on fairness in rankings to rankings recovered from pairwise comparisons. We define a respective group-conditioned error measure. We focus on a single, binary sensitive attribute, and notions of group fairness. We investigate sampling approaches motivated by representation, popularity, and self-selection biases. We evaluate the impact of state-of-the-art ranking recovery algorithms on accuracy and fairness of the recovered rankings, using synthetic and empirical data.

Our research provides three main insights: (i) Under random sampling, using GNNRank has little benefit over less sophisticated group-unaware recovery methods such as David's Score, which only need a fraction of the computational resources. (ii) Post-processing methods have the potential to improve both accuracy and fairness, but out of the methods we evaluated, only FA*IR achieves this. (iii) Both Fairness-Aware PageRank and GNNRank with FA*IR post-processing can be deployed to partially mitigate existing biases and to improving the overall accuracy of recovered rankings. GNNRank with FA*IR post-processing achieves the lowest overall error, but can suffer from increased error difference between the groups. Fairness-Aware PageRank creates fair rankings regarding exposure- and error-difference, but rankings recovered with this method will be less accurate if bias is little.

While our findings could be applied to more than two groups by considering each group separately against all others, future work should investigate multiple and multinary sensitive attributes, as well as different group sizes, other distributions for \textit{skill score} and \textit{bias}, and additional sampling approaches. Algorithms should also be evaluated against newly collected empirical data comprising both ground-truth scores and human annotated pairwise comparisons. Given the limitations of existing ranking recovery methods, we encourage the development of dedicated methods for fairness-aware ranking recovery. We provide a Python package under MIT license alongside our paper to facilitate replication and future work on fair ranking recovery\footnote{\url{https://github.com/wanLo/fairpair}}.






\bibliography{main}

\end{document}